\documentclass[11pt,a4paper]{article}

\pdfoutput=1

\usepackage{comment}
\usepackage{jcappub}

\usepackage{physics}
\usepackage{subfig}
\usepackage{float}
\usepackage{xcolor}
\usepackage{appendix}
\usepackage{booktabs}

\newcommand{\be}{\begin{equation}}
\newcommand{\ee}{\end{equation}}
\newcommand{\bea}{\begin{eqnarray}}

\newcommand{\eea}{\end{eqnarray}}

\newcommand{\pd}{\partial}
\newcommand{\MP}{M_\text{P}}

\usepackage{orcidlink}

\title{\centering Natural inflation in Palatini $F(R)$}

\author[a]{N. Bostan,}
\author[b]{R. H. Dejrah\orcidlink{0000-0002-8110-296X},}
\author[c,d]{C. Dioguardi,}
\author[d]{and A. Racioppi}

\affiliation[a]{Department of Physics, Faculty of Science, Marmara University, 34722 Istanbul, Türkiye}
\affiliation[b]{Department of Physics, Faculty of Science, Ankara University,  06100 Ankara, Türkiye}
\affiliation[c]{Tallinn University of Technology, Akadeemia tee 23, 12618 Tallinn, Estonia}
\affiliation[d]{National Institute of Chemical Physics and Biophysics, R\"avala 10, 10143 Tallinn, Estonia}

\emailAdd{nilay.bostan@marmara.edu.tr}
\emailAdd{rafid.dejrah@gmail.com}
\emailAdd{christian.dioguardi@kbfi.ee}
\emailAdd{antonio.racioppi@kbfi.ee}

\abstract{
$F(R)$ Palatini gravity provides a robust framework for constructing viable inflationary potentials. In this study, we examine natural inflation and show that its consistency with observational data can be restored when the model is embedded within $F(R)$ Palatini gravity, specifically for $F(R) = R + \alpha R^n$ with $7/4 \lesssim n \leq 2$. For completeness, we also demonstrate that models with $n > 2$ do not yield comparable improvements, achieving partial agreement with the data only in the limit $n \rightarrow 2$.
}
\keywords{natural inflation, Palatini gravity}
\usepackage{caption}
\begin{document}
\maketitle
%%%%%%%%%%%%%%%%%%%%%%%%%%%%%%%%%%%%%%%%%%%%%%%%%%%%%%%%%%%%%%%%%%%%%%%%%%%%%%%%%%%%%%%%
\section{Introduction}
The Standard Model of Cosmology is built on the cosmological principle, which asserts that the Universe is homogeneous and isotropic on sufficiently large scales. Observations of the Cosmic Microwave Background (CMB) \cite{Planck2018_cosmology}, large galaxy surveys \cite{SDSS2017,2dFGRS2001}, and the distribution of large-scale structures strongly support this principle, showing that the Universe is remarkably uniform across hundreds of megaparsecs. Nevertheless, explaining this uniformity, along with the near-flat geometry of the Universe and the absence of exotic relics, requires a mechanism operating in the very early Universe. The most compelling solution is a brief phase of rapid accelerated expansion~\cite{Starobinsky:1980te,Guth:1980zm,Linde:1981mu,Albrecht:1982wi}, commonly referred to as \emph{inflation}. In addition to resolving these issues, inflation provides a natural origin for the tiny primordial fluctuations that later evolve into galaxies, clusters, and the cosmic web.

A simple realization of inflation is through a scalar field, called the \emph{inflaton}, whose potential energy drives the accelerated expansion. The detailed shape of the inflaton potential determines both the duration of inflation and the properties of the resulting primordial perturbations. High-precision measurements of the CMB by \emph{Planck} and BICEP/\emph{Keck} \cite{Planck2018:inflation,BICEP:2021xfz} have ruled out the simplest minimally coupled models, motivating the exploration of more elaborate constructions. Among these, \emph{natural inflation} \cite{Freese:1990rb} stands out as a particularly elegant scenario: the inflaton is realized as a pseudo-Nambu-Goldstone boson with a periodic potential. This structure naturally keeps the potential flat, reducing the need for fine-tuning. Although theoretically appealing, the original natural inflation model is now in tension with observations, prompting numerous extensions and modifications to reconcile it with current data \cite{Kim:2004rp,Visinelli:2011jy,Achucarro:2015rfa,Ferreira:2018nav,Antoniadis:2018yfq,Salvio:2019wcp,Simeon:2020lkd,McDonough:2020gmn,Salvio:2021lka,Salvio:2022mld,Bostan:2022swq,Salvio:2023cry,Mukuno:2024yoa,Racioppi:2024zva,Lorenzoni:2024krn,Michelotti:2024bbc,Bostan:2025vkt}.

A promising approach is to embed inflation in the Palatini formulation of gravity (e.g. \cite{Koivisto:2005yc,Bauer:2008zj,Jarv:2020qqm,Gialamas:2023flv} and refs. therein), in which the metric and connection are treated as independent variables. This framework modifies the way the scalar field interacts with gravity, potentially altering the predictions for inflation, often reducing the required field excursion and relaxing observational constraints from the CMB. Within this context, $F(R)$ theories provide a versatile extension (e.g. \cite{Enckell:2018hmo,Antoniadis:2018ywb,Gialamas:2019nly,Gialamas:2020snr,Dimopoulos:2020pas,Karam:2021sno,Dioguardi:2021fmr,Dimopoulos:2022tvn,Dimopoulos:2022rdp,Dioguardi:2022oqu,Dioguardi:2023,Kuralkar:2025hoz,Dioguardi:2025mpp,Dioguardi:2025vci,Dimopoulos:2025fuq} and refs. therein), allowing the effective inflationary potential in the Einstein frame to be adjusted. In this work, we focus on models in which the potential remains bounded and positive, as in natural inflation, demonstrating that such scenarios can reconcile theory with observational data while preserving the desirable theoretical properties of the original model.

%%%%%%%%%%%%%%%%%%%%%%%%%%%%%%%%%%%%%%%%%%%%%%%%%%%%%%%%%%%%%%%%%%%%%%%%%%%%%%%%%%%%%%%%
%%%%%%%%%%%%%%%%%%%%%%%%%%%%%%%%%%%%%%%%%%%%%%%%%%%%%%%%%%%%%%%%%%%%%%%%%%%%%%%%%%%%%%%%
\section{Palatini $F(R)$ theories}\label{sec:Palatini}

Let us begin by considering the following action, minimally coupled to $F(R)$ gravity, where we set the reduced Planck mass $\MP \equiv 1$ and adopt a space-like metric signature:
\begin{equation} \label{eq:actionFR}
S_J = \int \dd^4 x \, \sqrt{-g_J} \left[ \frac{1}{2} F(R(\Gamma)) - \frac{1}{2} g_J^{\mu\nu} \partial_\mu \phi \, \partial_\nu \phi - V(\phi) \right] \, ,
\end{equation}
where $\phi$ is the inflaton scalar field. Here, $g_J^{\mu\nu}$ denotes the Jordan-frame metric, and $R(\Gamma) = g_J^{\mu\nu} R_{\mu\nu}(\Gamma)$ is the Ricci scalar obtained by contracting the metric-independent Ricci tensor with the Jordan-frame metric. This framework has been extensively studied in \cite{Dioguardi:2021fmr} for positive-definite potentials, and in \cite{Dioguardi:2022oqu} for well-defined negative potentials. A similar setup for more general functions of the form $F(R,X)$, with $X = - g_J^{\mu\nu} \partial_\mu \phi \, \partial_\nu \phi$, was considered in \cite{Dioguardi:2023}.
In this work, we focus on the complementary case in which the scalar potential is bounded from above, specifically considering natural inflation. 

As shown in \cite{Dioguardi:2021fmr}, the action \eqref{eq:actionFR} can be rewritten in the Einstein frame, where the gravitational action is linear in the Ricci scalar. As a first step, we introduce an auxiliary field $\zeta$ to rewrite the action as:
\begin{equation} \label{eq:action:zeta:J}
S_J = \int \dd^4 x \, \sqrt{-g_J} \left[ \frac{1}{2} \left( F(\zeta) + F'(\zeta) (R - \zeta) \right) - V(\phi) \right] \, ,
\end{equation}
where the prime denotes differentiation with respect to the argument. Next, we perform the conformal transformation:
\begin{equation}
g_{E\,\mu\nu} = F'(\zeta) \, g_{J\,\mu\nu} \, ,
\end{equation}
where $g_{E\,\mu\nu}$ is the Einstein-frame metric. Since the Ricci tensor is independent of the metric in the Palatini formulation, this transformation leads to:

\begin{equation} \label{eq:action:zeta:E}
S_E = \int \dd^4 x \, \sqrt{-g_E} \left[ \frac{1}{2} R_E - \frac{1}{2} g_E^{\mu\nu} \partial_\mu \chi \, \partial_\nu \chi - U(\chi, \zeta) \right] \, ,
\end{equation}
where we have defined a canonically normalized scalar field:
\begin{equation} \label{eq:dchidphi}
\frac{\pd \chi}{\pd \phi} = \sqrt{\frac{1}{F'(\zeta)}} \, ,
\end{equation}
and an effective scalar potential:
\begin{equation} \label{eq:Uchizeta}
U(\chi, \zeta) = \frac{V(\phi(\chi))}{F'(\zeta)^2} - \frac{F(\zeta)}{2 F'(\zeta)^2} + \frac{\zeta}{2 F'(\zeta)} \, .
\end{equation}
The equation of motion for the auxiliary field is obtained by varying \eqref{eq:action:zeta:E} with respect to $\zeta$, which yields:
\begin{equation} \label{eq:EoMzetafull}
2 F(\zeta) - \zeta F'(\zeta) -  \, \partial^\mu \phi \partial_\mu \phi \, F'(\zeta) - 4 V(\phi) = 0 \, .
\end{equation}
In general, this equation cannot be solved explicitly for $\zeta$. However, for the computation of inflationary observables, we can adopt the slow-roll approximation, in which the energy of the scalar field is dominated by its potential. Under this approximation, Eq. \eqref{eq:EoMzetafull} reduces to:
\begin{equation} \label{eq:EoMzeta}
G(\zeta) = V(\phi) \, ,
\end{equation}
with
\begin{equation} \label{eq:G}
G(\zeta) \equiv \frac{1}{4} \left[ 2 F(\zeta) - \zeta F'(\zeta) \right] \, .
\end{equation}
Although it is generally not possible to find an explicit solution for $\zeta(\phi)$, one can use \eqref{eq:EoMzeta} in \eqref{eq:Uchizeta} to express the Einstein-frame potential as function of just $\zeta$:
\begin{equation} \label{eq:Uzeta}
U(\zeta) = \frac{\zeta}{4 F'(\zeta)} \, .
\end{equation}
The slow-roll parameters can then be computed as derivatives of $U(\zeta)$ with respect to the canonical field $\chi$:
\begin{equation} \label{eq:epsilon_zeta}
\epsilon(\zeta) = \frac{1}{2} \left( \frac{\partial U / \partial \chi}{U} \right)^2 = \frac{1}{2} g^2 \left( \frac{U'}{U} \right)^2 \, ,
\end{equation}
\begin{equation} \label{eq:eta_zeta}
\eta(\zeta) = \frac{\partial^2 U / \partial \chi^2}{U} = \frac{g g' U' + g^2 U''}{U} \, ,
\end{equation}
where

\begin{equation} \label{eq:gzeta}
g(\zeta) \equiv \frac{\partial \zeta}{\partial \chi} = \sqrt{F'(\zeta)} \left( \frac{\pd G}{\pd \zeta} \frac{\pd V^{-1}}{\pd G} \right)^{-1} \, .
\end{equation}
The CMB observables are then given by:
\begin{align}
r(\zeta) &= 16 \, \epsilon(\zeta) = 8 \, g^2 \left( \frac{U'}{U} \right)^2 , \label{eq:r_zeta} \\
n_s(\zeta) &= 1 + 2 \, \eta(\zeta) - 6 \, \epsilon(\zeta) 
= 1 + \frac{2 g}{U^2} \left( g' U' U + g U'' U - 24 g U'^2 \right) , \label{eq:ns_zeta} \\
A_s(\zeta) &= \frac{U}{24 \pi^2 \epsilon(\zeta)} = \frac{U^3}{12 \pi^2 g^2 U'^2} \, . \label{eq:As_zeta}
\end{align}
Finally, the number of $e$-folds of remaining inflation can be expressed as follows:
\begin{equation} \label{eq:N_zeta}
N_e = \int_{\chi_{\rm end}}^{\chi_N} \frac{U}{\partial U / \partial \chi} \, \dd \chi
= \int_{\zeta_{\rm end}}^{\zeta_N} \frac{U}{g^2 \, \partial U / \partial \zeta} \, \dd \zeta \, ,
\end{equation}
where $\zeta_{\rm end}$ is defined by $\epsilon(\zeta_{\rm end}) = 1$, and $\zeta_N$ corresponds to the field value at which the pivot scale exits the horizon. Further details on this formalism can be found in \cite{Dioguardi:2021fmr}.

In this work, we consider as the Jordan-frame potential, the natural inflation one:
\begin{equation}
V(\phi) = \Lambda^4 \left[ 1 + \cos\left( \frac{\phi}{M} \right) \right] ,
\end{equation}
together with a polynomial $F(R)$ function of the form:
\begin{equation} \label{eq:F(R)}
F(R) = R + \alpha R^n \, .
\end{equation}
Due to the distinct properties of $G(\zeta)$, we analyze the cases $n \leq 2$ and $n > 2$ in separate sections.

%%%%%%%%%%%%%%%%%%%%%%%%%%%%%%%%%%%%%%%%%%%%%%%%%%%%%%%%%%%%%%%%%%%%%%%%%%%%%%%%%%%%%%%%
\begin{figure}[t]
    \centering
    \includegraphics[width=0.45\textwidth]{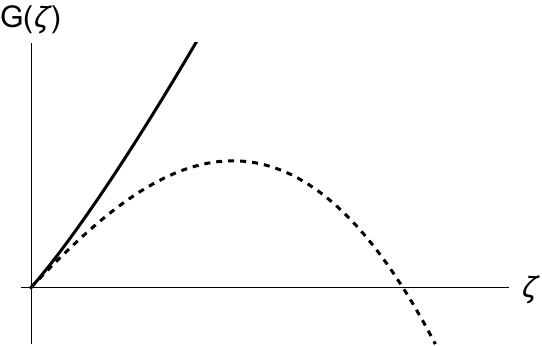}
    \qquad
    \includegraphics[width=0.45\textwidth]{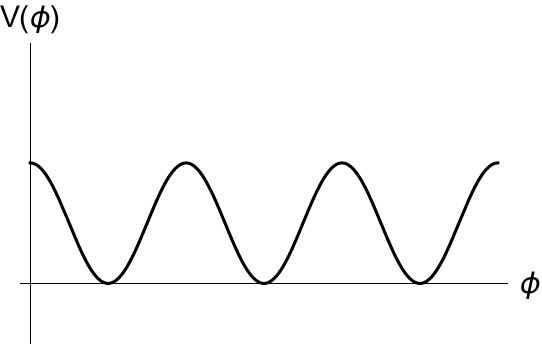}
    \caption{Reference plots of $G(\zeta)$ (left), generated using $F(R) = R + \alpha R^n$ for $n \leq 2$ (continuous) and $n>2$ (dashed), and the natural inflation potential $V(\phi)$ (right). Note that, for $n>2$ the condition $G(\zeta) = V(\phi)$ can be satisfied for any value of $\phi$, provided that the local maximum of $G$ exceeds the maximum of $V$.}
    \label{fig:GvsV}
\end{figure}

%%%%%%%%%%%%%%%%%%%%%%%%%%%%%%%%%%%%%%%%%%%%%%%%%%%%%%%%%%%%%%%%%%%%%%%%%%%%%%%%%%%%%%%%

%%%%%%%%%%%%%%%%%%%%%%%%%%%%%%%%%%%%%%%%%%%%%%%%%%%%%%%%%%%%%%%%%%%%%%%%%%%%%%%%%%%%%%%%
\section{Natural inflation for $n\leq 2$}\label{sec:n<2}
\begin{figure}[t!]%
    \centering
    \subfloat[]{\includegraphics[width=0.45\textwidth]{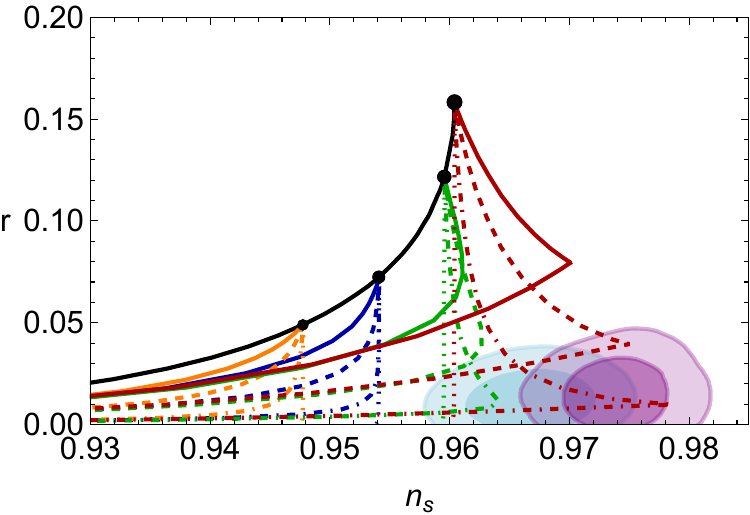}}%
    \qquad
    \subfloat[]{\includegraphics[width=0.46\textwidth]{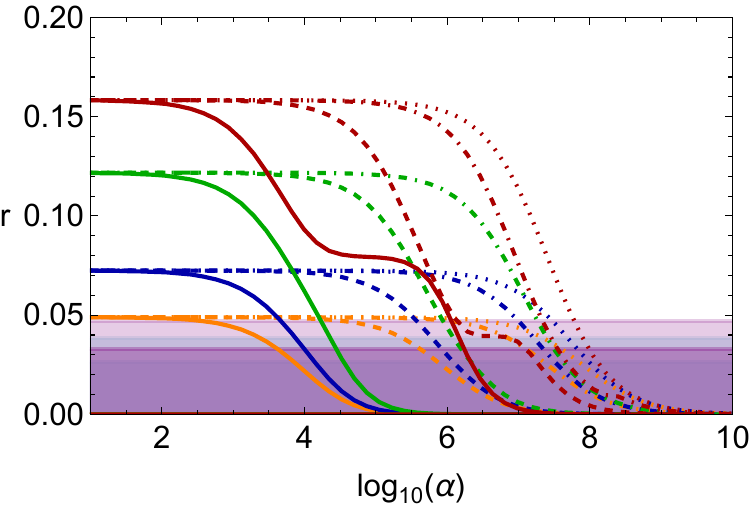}}%

    \subfloat[]{\includegraphics[width=0.455\textwidth]{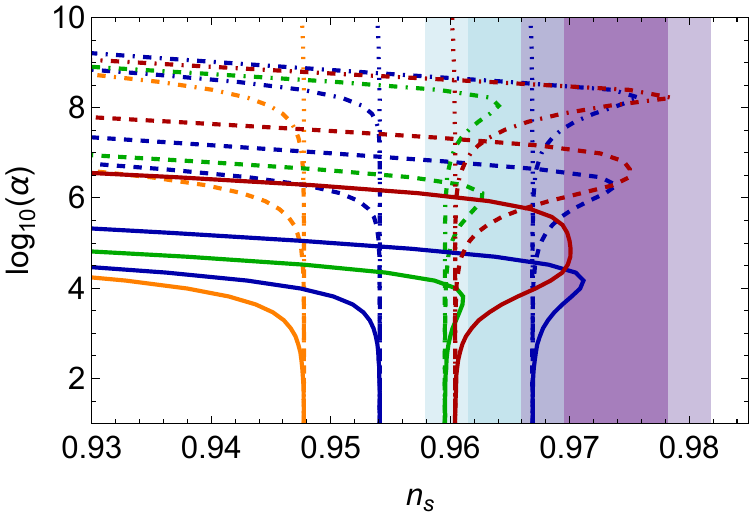}}%
    \qquad
    \subfloat[]{\includegraphics[width=0.475\textwidth]{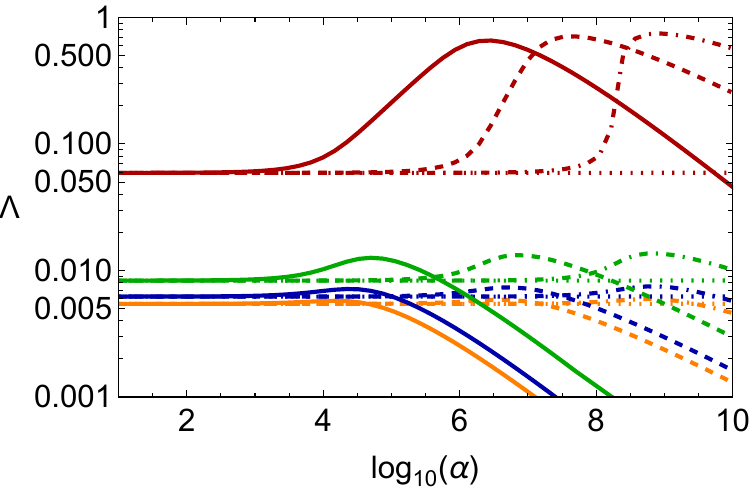}}
   \caption{(a) $r$ vs. $n_s$, (b) $r$ vs. $\log_{10}(\alpha)$, (c) $n_s$ vs. $\log_{10}(\alpha)$, and (d) $\Lambda$ vs. $\log_{10}(\alpha)$ for $n \leq 2$ and $N_e=50$ with $n = 3/2$ (continuous), $n = 7/4$ (dashed), $n = 31/16$ (dot-dashed), and $n = 2$ (dotted), and with $M = 5$ (orange), $M = 6$ (blue), $M = 10$ (green), and $M = 500$ (red). The black line indicates the original prediction for natural inflation, while the dots correspond to predictions for fixed $M$, with larger dots representing increasing $M$. Contours display the 68\% and 95\% confidence levels based on the latest combinations from BICEP/\emph{Keck}\cite{BICEP:2021xfz} (cyan), and ACT collaborations \cite{ACT:2025tim} (purple).}
    \label{fig:n_less_2_N50}
\end{figure}
\begin{figure}[t!]%
    \centering
    \subfloat[]{\includegraphics[width=0.45\textwidth]{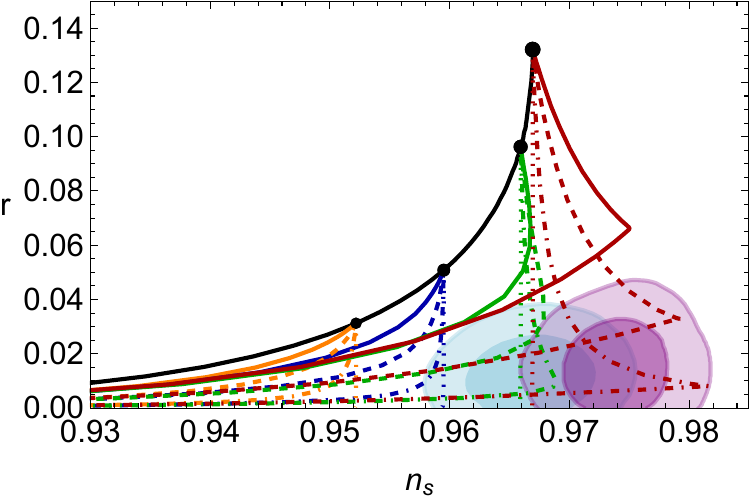}}%
    \qquad
    \subfloat[]{\includegraphics[width=0.46\textwidth]{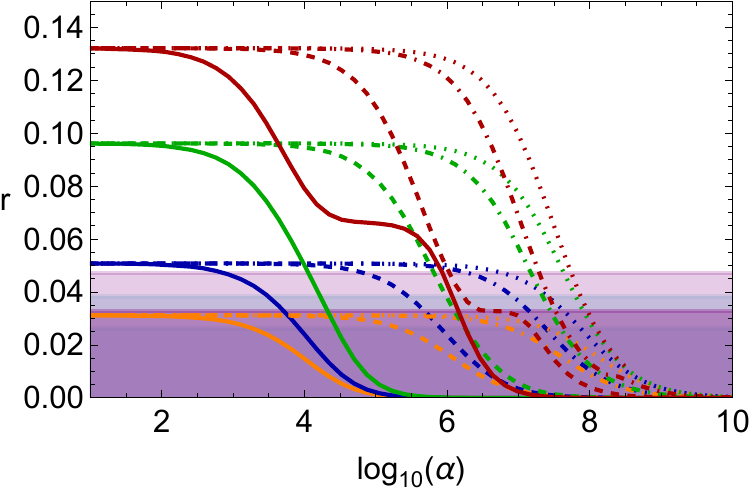}}%
    
    \subfloat[]{\includegraphics[width=0.455\textwidth]{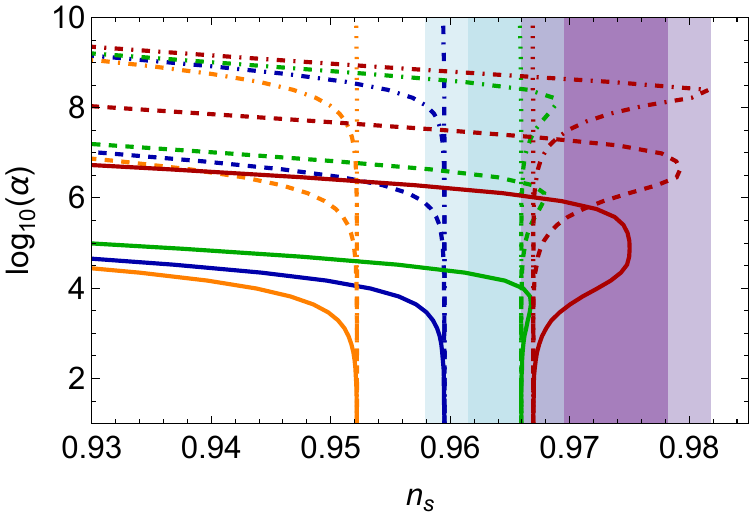}}%
    \qquad
    \subfloat[]{\includegraphics[width=0.475\textwidth]{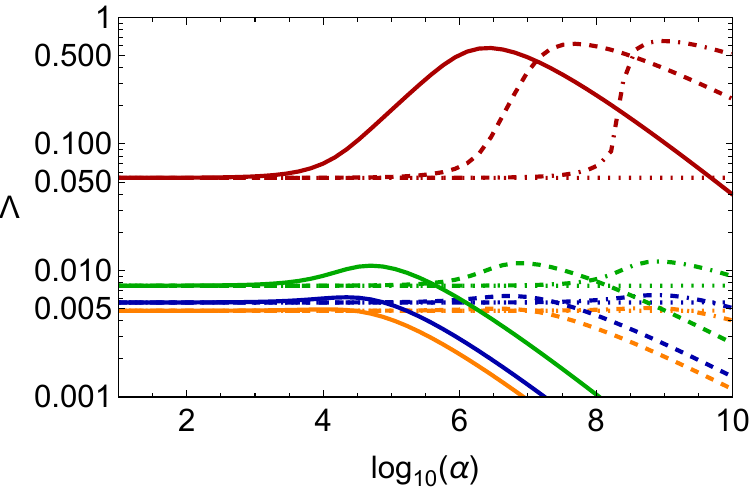}}%

   \caption{Same as Fig.~\ref{fig:n_less_2_N50}, but for $N_e = 60$.}
    \label{fig:n_less_2}
\end{figure}
For the class of $F(R)$ models defined in Eq.~\eqref{eq:F(R)}, we obtain from Eq.~\eqref{eq:G} that:
\begin{equation}
G(\zeta) = \frac{1}{4} \left(\zeta + \alpha (2-n) \zeta^n \right),
\end{equation}
which is positive and monotonically increasing for $\zeta > 0$, provided that $1 < n \leq 2$. We do not consider the case $n<1$ as we want that $F(R)\sim R$ as $R\rightarrow 0$, in order to recover General Relativity in the low energy limit. The case $n=2$ was already studied explicitly in \cite{Enckell:2018hmo,Antoniadis:2018yfq}. Therefore, we focus on the range $1 < n < 2$. To obtain expressions for the CMB observables, we adopt the formalism derived in \cite{Dioguardi:2021fmr} and briefly outlined in Section~\ref{sec:Palatini}. In terms of $\zeta_N$, we then have:

\begin{align} \label{eq:CMB_observables_1}
n_s(\zeta_N) &= 1 - \frac{16 \Lambda^4 - \zeta_N}{M^2 \zeta_N G'(\zeta_N)} 
- \frac{2 \alpha (2-n) \left[ 4 \Lambda^4 (3-n) - \zeta_N (2-n) \right] \zeta_N^{n-2}}{M^2 G'(\zeta_N)} \\
&\quad - \frac{\alpha^2 (n-2)^2 (2n-3) \zeta_N^{2n-2}}{M^2 G'(\zeta_N)}, \nonumber \\
r(\zeta_N) &= \frac{128 G(\zeta_N) \left( 2 \Lambda^4 - G(\zeta_N) \right)}{ M^2 \zeta_N^2 F'(\zeta_N)}, \\
A_s(\zeta_N) &= \frac{ M^2 \zeta_N^3}{768 \pi^2 G(\zeta_N) \left( 2 \Lambda^4 - G(\zeta_N) \right)}, \\
N_e(\zeta_N) &= \int^{\zeta_N}_{\zeta_{\rm end}} \frac{M^2 \zeta G'(\zeta)}{4 G(\zeta) \left( 2 \Lambda^4 - G(\zeta) \right)} d\zeta.
\label{eq:CMB_observables_4}
\end{align}
The numerical results for the $n \leq 2$ scenario are shown in Figs.~\ref{fig:n_less_2_N50} and \ref{fig:n_less_2}, corresponding to $N_e = 50$ and $N_e = 60$, respectively. Panel (a) shows $r$ vs.\ $n_s$, panel (b) shows $r$ vs.\ $\log_{10}(\alpha)$, panel (c) shows $n_s$ vs.\ $\log_{10}(\alpha)$, and panel (d) shows $\Lambda$ vs.\ $\log_{10}(\alpha)$, for $n = 3/2$ (continuous), $n = 7/4$ (dashed), $n = 31/16$ (dot-dashed), and $n = 2$ (dotted), with $M = 5$ (orange), $M = 6$ (blue), $M = 10$ (green), and $M = 500$ (red). The black line denotes the original prediction of natural inflation, while the dots show the predictions for fixed $M$; larger dots correspond to increasing values of $M$.
The amplitude of the power spectrum, $A_s$, is fixed to its observed value $A_s \simeq 2.1 \cdot 10^{-9}$ \cite{Planck2018:inflation}. Contours indicate the 68\% and 95\% confidence levels based on the latest combinations from BICEP/\emph{Keck}~\cite{BICEP:2021xfz} (cyan), and ACT collaborations \cite{ACT:2025tim} (purple).
All plots were obtained by fixing $(M,n)$ and varying $\alpha$ in the range $0 < \alpha < 10^{10}$. As it is well known, the shift in $N_e$ does not alter the dependence on the free parameters of the model\footnote{Larger values of $N_e$ probe field values farther from the end of inflation, where the potential is flatter, naturally leading to smaller $\epsilon$ and, consequently, lower $r$ and higher $n_s$, while leaving the dependence on $M$, $\alpha$, and $n$ unchanged.}, and therefore the following discussion applies to both Figs.~\ref{fig:n_less_2_N50} and \ref{fig:n_less_2}.

\begin{figure}[t!]%
\hspace{-0.5cm}
    \includegraphics[width=1.06\textwidth]{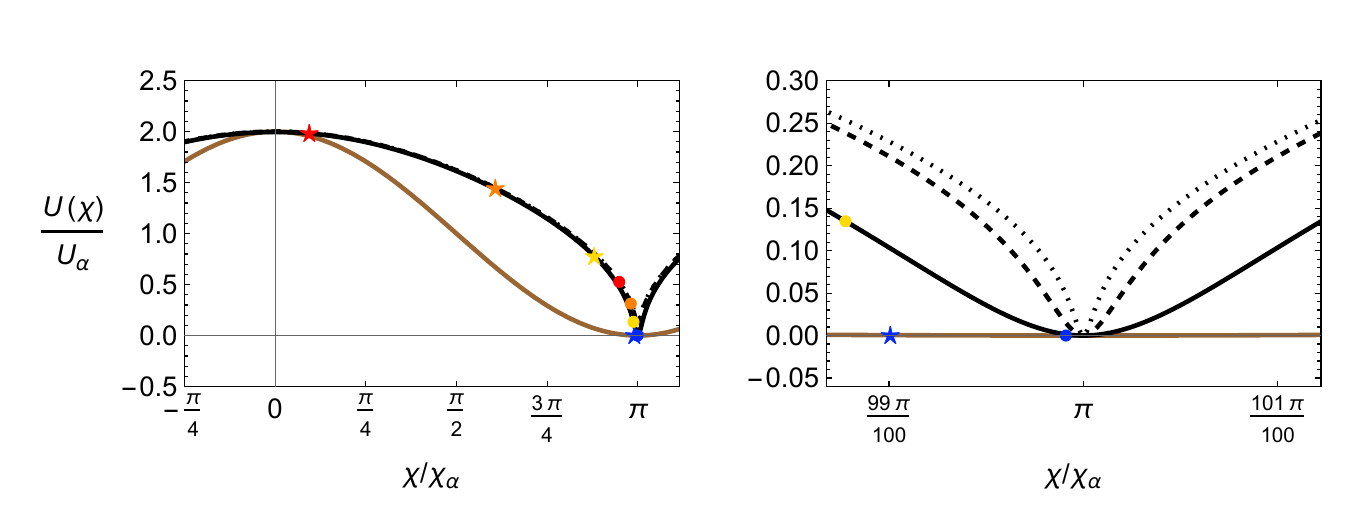}
   \caption{The Einstein-frame potential $U(\chi)$ for $\alpha = 0$ (brown), $\alpha = 3.16 \cdot 10^6$ (black, continuous), $\alpha = 10^7$ (black, dashed), and $\alpha = 10^8$ (black, dotted), with $M = 500$ and $n = 7/4$ (left), along with a zoomed-in view of the same potential (right). We indicate $\chi_N$ (stars) and $\chi_{\rm end}$ (dots) for $\alpha = 0$ (blue), $3.16 \cdot 10^6$ (yellow), $10^7$ (orange), and $10^8$ (red). The value $\alpha = 3.16 \cdot 10^6$ corresponds to the maximum of $n_s$, i.e., the ``knee'' in Fig.~\ref{fig:n_less_2}. The potentials are normalized by $U_{\alpha}$ (defined in Eq.~\eqref{eq:U_max}) and the Einstein-frame canonical field $\chi$ by $\chi_\alpha$ (defined in Eq.~\eqref{eq:chimax}), so that the potential appears with a period of $2\pi$ in the plot. Further details are given in the text.}

    \label{fig:U_plot}
\end{figure}

In panel (a), we observe that as $n$ approaches $2$, the predictions move towards lower values of $r$, while $n_s$ spans a larger range. Panel (b) shows that the tensor-to-scalar ratio $r$ decreases for increasing $\alpha$, similar to the behavior observed for positive unbounded potentials in \cite{Dioguardi:2021fmr}. In panel (c), for $M \gtrsim 10$, $n_s$ increases until reaching a maximum value that depends on both $M$ and $n$, and then decreases towards values disfavored by both combinations. This is reflected in panel (b) by a temporary plateau in $r$, most noticeable for $M = 500$ and $n=3/2,7/4$. 
Such behavior corresponds to the ``knee'' shape of the $r$ vs.\ $n_s$ predictions in panel (a), which can be understood as follows. At very large $M$, natural inflation approximates quadratic inflation. One would therefore naively expect that for $M = 500$, the results of \cite{Dioguardi:2021fmr} are reproduced. Indeed, this holds from $\alpha = 0$ up to the maximum of $n_s$ (the corresponding value for $\alpha$ depends on $n$). The ``knee'' is located precisely in the strong coupling limit studied in \cite{Dioguardi:2021fmr}:
\bea
  r(n) &\simeq& \frac{8 (2-n)}{N_e} \, \label{eq:r:limit},\\
n_s(n) &\simeq& 1 - \frac{3-n}{N_e}, \, \label{eq:ns:limit} 
\eea
where we remind the reader that $1 < n < 2$. These predictions correspond precisely to those of monomial inflation with power $\bar{n} = 2(2 - n)$ in the large-$N_e$ approximation. Therefore, when the ``knee'' is reached, the model effectively behaves like $\phi^{\bar{n}}$ inflation. 
As $\alpha$ increases, the predictions remain approximately stable around this configuration, but $\zeta_N$ (and consequently $\phi_N$ and $\chi_N$) slowly grows. Eventually, the shape of the potential in the inflationary region ceases to resemble a monomial form and becomes hilltop-like, explaining the subsequent turn towards lower values of $n_s$. To visualize all this in terms of the scalar field potential, we show in Fig.~\ref{fig:U_plot} the Einstein-frame potential $U(\chi)$ for $\alpha = 0$ (brown), $\alpha = 3.16 \cdot 10^6$ (black, continuous), $\alpha = 10^7$ (black, dashed), and $\alpha = 10^8$ (black, dotted), with $M = 500$ and $n = 7/4$ (left), alongside a zoomed-in view of the same potential  around the minimum (right). Notice that in Fig. \ref{fig:U_plot} the potentials are normalized by:
\be\label{eq:U_max}
U_{\alpha}\equiv 2 
U(\zeta_{max}),
\ee
with $\zeta_{\rm max}$ defined as the value of $\zeta$ satisfying:
\begin{equation}
2 \Lambda^4 = G(\zeta) \, .
\end{equation}
Moreover, the Einstein-frame canonical field $\chi$ is normalized by $\chi_\alpha$, defined as:
\begin{equation}\label{eq:chimax}
\chi_\alpha = \frac{1}{\pi} \int_0^{\zeta_{\rm max}} \frac{1}{g(\zeta)} \, d\zeta,
\end{equation}
so that it appears with a period of $2\pi$ in the plot. The subscript $\alpha$ emphasizes that $U_\alpha$ and $\chi_\alpha$ depend on $\alpha$.
 We indicate $\chi_N$ (stars) and $\chi_{\rm end}$ (dots) for $\alpha = 0$ (blue), $3.16 \cdot 10^6$ (yellow), $10^7$ (orange), and $10^8$ (red). The value $\alpha = 3.16 \cdot 10^6$ corresponds to the maximum of $n_s$, i.e., the ``knee'' in Fig.~\ref{fig:n_less_2}. For sufficiently small $\alpha$\footnote{It can be computed numerically that for $n = 7/4$ and $M = 500$, the higher-order contribution becomes relevant for $\alpha \gtrsim 3 \cdot 10^5$ (see Figs.~\ref{fig:n_less_2_N50} and \ref{fig:n_less_2}).} the higher-order term is negligible, and the Einstein-frame potential coincides with the Jordan-frame potential. Once the higher-order term starts dominating, up to $\alpha \lesssim 3.16 \cdot 10^6$, the Einstein-frame potential can be approximated by a monomial potential with $\bar{n} = 2(2-n)$ in the inflationary region.  For larger $\alpha$, this approximation breaks down, causing $n_s$ to decrease and producing the ``tail'' towards lower scalar spectral index values observed in panel (a) of Fig.~\ref{fig:n_less_2_N50} and \ref{fig:n_less_2}.

In panel (d), the predictions for $\Lambda$ as a function of $\log_{10}(\alpha)$ are obtained by imposing $A_s \simeq 2.1 \cdot 10^{-9}$. The behavior of $\Lambda$ directly reflects the ``knee'' structure seen in panel (a). it initially increases up to a maximum value:
 \be
 \Lambda \simeq M^\frac{1}{2} \left[ \frac{3 \pi^2 (2-n) A_s \alpha }{2^{2-3 n} \left(\frac{n}{N_e}\right)^{n-3} } \right]^\frac{1}{4 (2-n)}\, ,
 \label{eq:L:limit}
 \ee
which corresponds to the $\phi^{\bar n}$ inflation limit, after which it decreases as $\alpha$ increases.\\

The predicted values of $r$ tend to be relatively high, with only scenarios featuring both large $\alpha$ ($\gtrsim 10^6$) and large $M$ ($M \ge 10$ for $N_e = 50$ and $M \ge 6$ for $N_e = 60$ in our ensemble) being consistent with the cosmological data. Depending on the specific parameter values, the predictions can agree with the constraints from the BICEP/\emph{Keck} collaboration, the ACT collaboration, or both. 

For $N_e = 50$, compatibility with the ACT combination at $\lesssim 2\sigma$ is achieved for $M > 10$ ($M = 500$ in our ensemble) and $n \gtrsim 7/4$, within an $\alpha$ range that depends on $n$, roughly $10^6$--$10^8$. Compatibility with the BICEP/\emph{Keck} combination at $\lesssim 2\sigma$ is obtained already for lower values, $M \gtrsim 10$, provided $n \gtrsim 7/4$. Simultaneous compatibility with both combinations is achieved for $n \gtrsim 7/4$, $M > 10$ ($M = 500$ in our ensemble), and $\alpha \sim 10^7$–$10^8$.\\
Increasing the number of $e$-folds to $N_e = 60$ (Fig.~\ref{fig:n_less_2}) shifts the predictions toward lower $r$ and slightly higher $n_s$, allowing the $M = 6$ predictions to fall within the 95\% confidence regions of the BICEP/\emph{Keck} combination. Compatibility with the ACT combination at $\lesssim 2\sigma$ is achieved for $M \gtrsim 10$ and $n \gtrsim 7/4$, within an $n$-dependent range of $\alpha$, roughly between $10^6$ and $10^8$. Agreement with both combinations is realized for $n \gtrsim 7/4$, $10 < M < 500$, and $\alpha \sim 10^7$--$10^8$.
Our analysis for $1<n\leq2$ shows that inflationary scenarios consistent with current observations are obtained only for $n$ values quite close to $2$. The parameter $n$ controls how strongly the higher-order curvature term flattens the Einstein frame scalar potential $U$, and therefore suppresses tensor-to-scalar ratio $r$. As $n$ approaches $2$ from below (and $\alpha$ increases), the potential progressively acquires a plateau-like profile (see Fig. \ref{fig:U_plot}), similar to that found in $R^2$ inflation, and leads to predictions that are consistent with current CMB data.
\section{Natural inflation for $n > 2$}\label{sec:n>2}
\begin{figure}[t!]%
    \centering
    \subfloat[]{\includegraphics[width=0.452\textwidth]{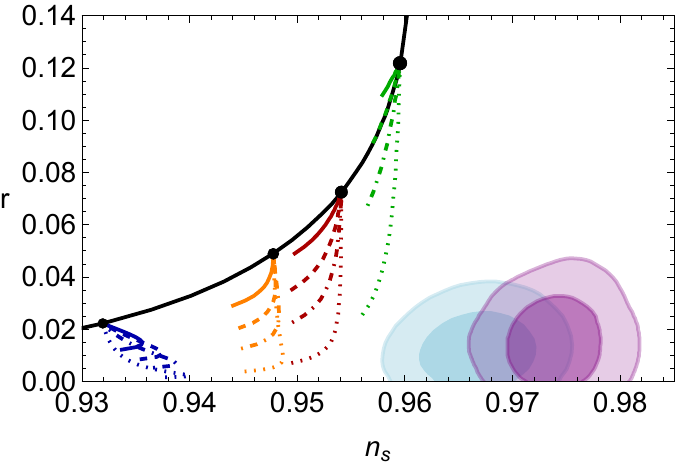}}%
    \qquad
    \subfloat[]{\includegraphics[width=0.45\textwidth]{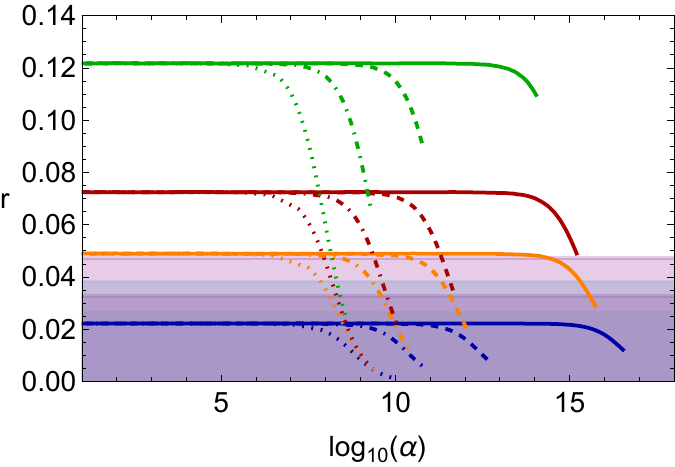}}%
    
    \subfloat[]{\includegraphics[width=0.45\textwidth]{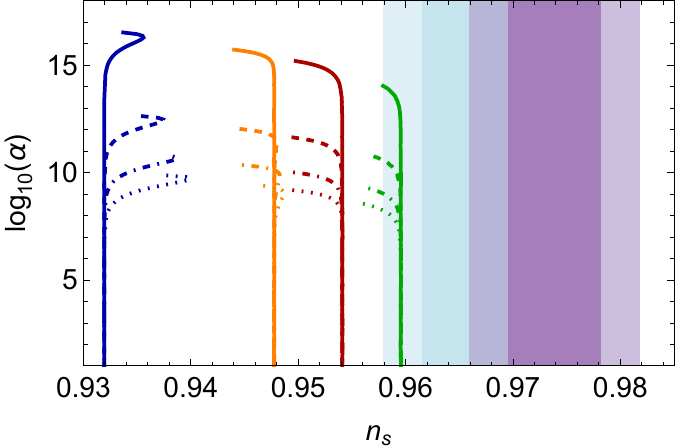}}%
    \qquad
    \subfloat[]{\includegraphics[width=0.455\textwidth]{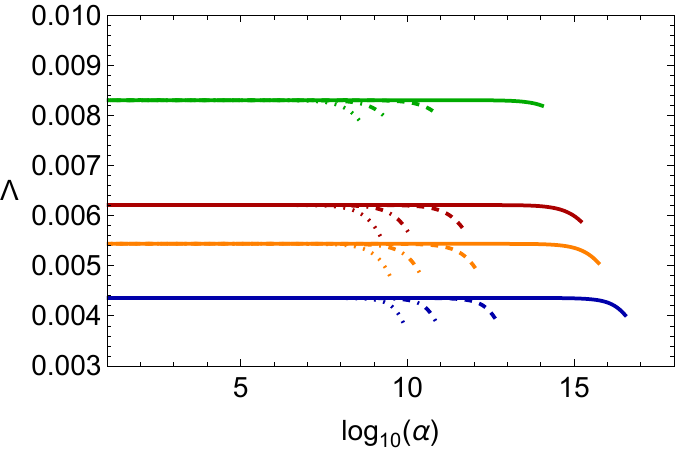}}%

   \caption{(a) $r$ vs. $n_s$, (b) $r$ vs. $\log_{10}(\alpha)$, (c) $n_s$ vs. $\log_{10}(\alpha)$, and (d) $\Lambda$ vs. $\log_{10}(\alpha)$ for $n > 2$ and $N_e = 50$, with $M = 4$ (blue), $M = 5$ (orange), $M = 6$ (red), and $M = 10$ (green), and $n = 3$ (continuous), $n = 5/2$ (dashed), $n = 9/4$ (dot-dashed), and $n = 33/16$ (dotted). Contours indicate the 68\% and 95\% confidence levels based on the latest combinations from the BICEP/\emph{Keck} \cite{BICEP:2021xfz} (cyan) and ACT \cite{ACT:2025tim} (purple) collaborations.}
    \label{fig:n_bigger_2_N50}
\end{figure}

\begin{figure}[t!]%
    \centering
    \subfloat[]{\includegraphics[width=0.452\textwidth]{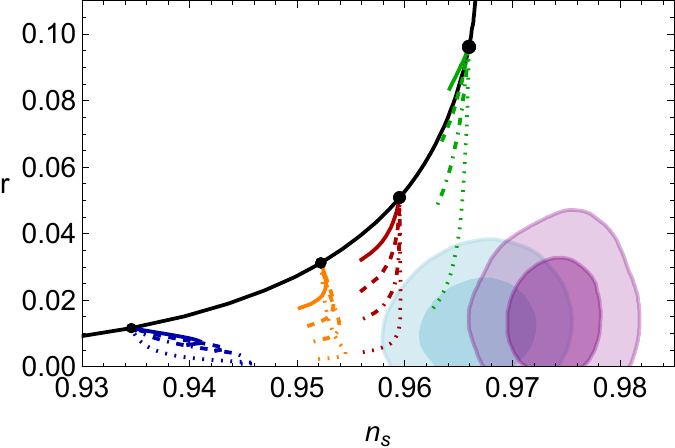}}%
    \qquad
    \subfloat[]{\includegraphics[width=0.45\textwidth]{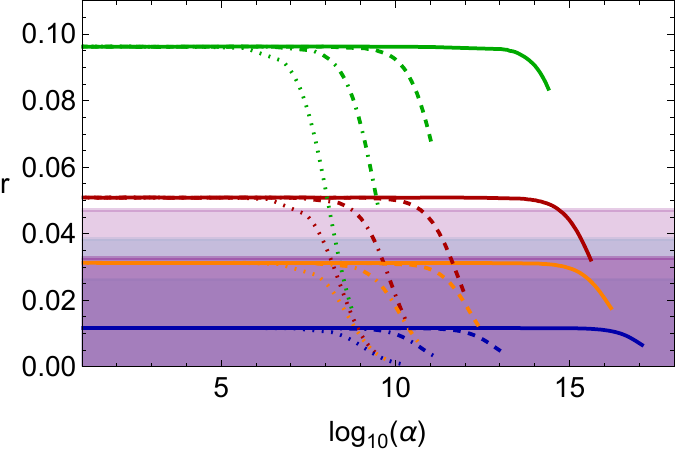}}%
    
    \subfloat[]{\includegraphics[width=0.45\textwidth]{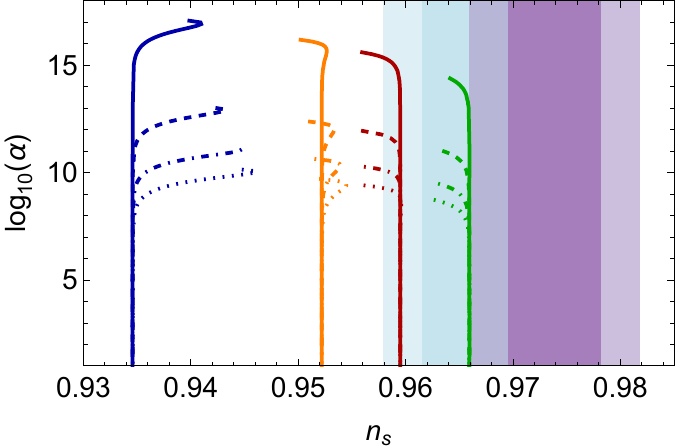}}%
    \qquad
    \subfloat[]{\includegraphics[width=0.455\textwidth]{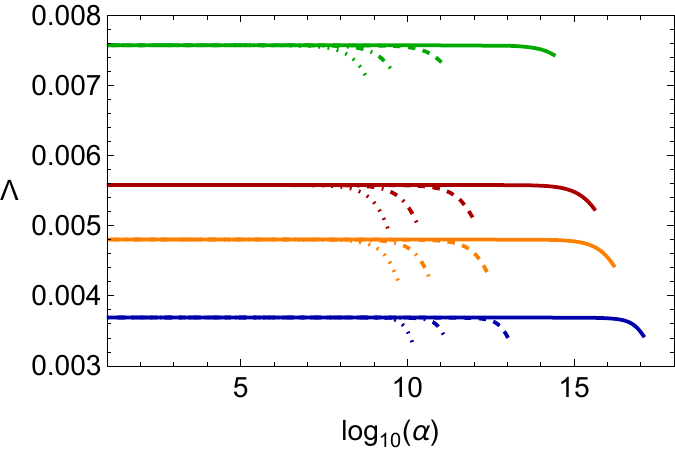}}%

   \caption{Same as Fig.~\ref{fig:n_bigger_2_N50}, but for $N_e = 60$.}
    \label{fig:n_bigger_2}
\end{figure}
We now consider the case with $n > 2$. The CMB observables are still given by Eqs.~\eqref{eq:CMB_observables_1} and \eqref{eq:CMB_observables_4}. However, a crucial difference arises: the function $G(\zeta)$ is now positive only in the range:
\begin{equation}
0 < \zeta < \zeta_0,
\end{equation}
with $\zeta_0 \equiv (\alpha(n-2))^{\frac{1}{1-n}}$.
At the same time, the function reaches its maximum at:
\begin{equation}
\zeta_M = (\alpha (n-2) n)^{\frac{1}{1-n}}, \label{eq:zetamax}
\end{equation}
for which:
\begin{equation}
G(\zeta_M) = \frac{n-1}{4n} \, \zeta_M. \label{eq:Gzetamax}
\end{equation}
As it can be understood from Fig.~\ref{fig:GvsV}, given the shape of $G(\zeta)$ we have two possibilities for mapping $\zeta$ to $\phi$. The first is to consider the range $0<\zeta<\zeta_M$, while the other is to consider $\zeta_M<\zeta<\zeta_0$. However, in the latter range it can be shown numerically that we always have $\epsilon(\zeta) \ll 1$ which doesn't allow a graceful exit from inflation. For this reason, in what follows we only consider the $0<\zeta<\zeta_M$ range.
Moreover, we need to require $G(\zeta_M) \geq 2 \, \Lambda^4$ in order to ensure that the mapping between the Jordan-frame and Einstein-frame potentials is well defined. This represents the main difference with the $n<2$ case. From Eqs.~\eqref{eq:zetamax} and \eqref{eq:Gzetamax}, we see that this condition cannot be satisfied for arbitrary values of $\alpha$, as $G(\zeta_M)$ decreases with increasing $\alpha$. The maximum allowed value of $\alpha$ can therefore be computed as:

\begin{equation}
\alpha_M = \frac{1}{n(n-2)} \left(\frac{n-1}{8 n \Lambda^4}\right)^{n-1}.
\end{equation}

Figures~\ref{fig:n_bigger_2_N50} and \ref{fig:n_bigger_2} show the results, respectively for $N_e = 50$ and $N_e = 60$, for (a) $r$ vs. $n_s$, (b) $r$ vs. $\log_{10}(\alpha)$, (c) $n_s$ vs. $\log_{10}(\alpha)$, and (d) $\Lambda$ vs. $\log_{10}(\alpha)$, with $M = 4$ (blue), $M = 5$ (orange), $M = 6$ (red), and $M = 10$ (green), and $n = 3$ (continuous), $n = 5/2$ (dashed), $n = 9/4$ (dot-dashed), and $n = 33/16$ (dotted). The amplitude of the power spectrum, $A_s$, is fixed to its observed value, $A_s \simeq 2.1 \cdot 10^{-9}$ \cite{Planck2018:inflation}. Contours indicate the 68\% and 95\% confidence levels based on the latest combinations from the BICEP/\emph{Keck} \cite{BICEP:2021xfz} (cyan) and ACT~\cite{ACT:2025tim} (purple) collaborations.

We observe that, since $\alpha$ cannot increase indefinitely, the predictions for $r$ vs. $n_s$ remain largely ruled out as $n \rightarrow 2$. This behavior is qualitatively similar to that reported in~\cite{Bostan:2025vkt} for the class of $F(R,X)$ theories, although the quantitative predictions for the CMB observables differ. Partial agreement with the BICEP/\emph{Keck} combination can only be achieved for $n \rightarrow 2$, with $M \gtrsim 6$ at $N_e = 60$ and $\alpha \gtrsim 10^8$. We therefore conclude that natural inflation can be rescued when embedded in $F(R)_{>2}$ Palatini gravity, but only within a restricted region of the parameter space.

\section{Conclusions}\label{sec:conclusion}

We have analyzed natural inflation within the Palatini $F(R)$ framework, considering $F(R) = R + \alpha R^n$. Our results show that for $n < 2$, the model can improve upon the predictions of standard natural inflation, particularly for $M \gtrsim 10$ and $7/4 \lesssim n \leq 2$, with $\alpha$ lying within an appropriate $n$-dependent range. In this regime, the predictions for $r$ and $n_s$ can satisfy the constraints from the BICEP/\emph{Keck} and/or ACT combinations. For $n > 2$, however, the allowed values of $\alpha$ are constrained by the observed scalar amplitude $A_s = 2.1 \cdot 10^{-9}$ and by the consistency of the equation of motion of the auxiliary field (Eq.~\ref{eq:EoMzeta}), which limits the contribution of the higher-order term to the inflationary dynamics. Consequently, $r$ remains relatively high and $n_s$ does not shift sufficiently, preventing full agreement with observational constraints. Partial agreement can only be achieved for $n \rightarrow 2$, if $M \gtrsim 6$ and $\alpha \gtrsim 10^8$. \\
Clearly, the $n<2$ is strongly favored over the $n>2$ case. However, the $n<2$ scenario still requires relatively tuned values for $n$ which are less natural choices than just $n=2$. On the other hand, with $7/4 < n < 2$, larger $n_s$ area can be covered at a given $M$, by varying $\alpha$. 

An additional implication of our analysis is that a viable inflationary phase occurs only if the axion decay constant is transPlanckian. This requirement is a well known shortcoming of natural inflation scenarios, which cannot be solved by our construction. Realizing an effective super-Planckian scale ordinarily demands additional structure, such as alignment mechanisms, kinetic mixing effects, or multiple axion fields (see e.g. \cite{Pajer:2013fsa} and refs therein). Such a completion is beyond the purpose of our work and reserve a comprehensive ultraviolet embedding for future analyses.

These results highlight that Palatini $F(R)$ corrections can resolve the compatibility issues of natural inflation with observations, but only within specific parameter ranges.

%%%%%%%%%%%%%%%%%%%%%%%%%%%%%%%%%%%%%%%%%%%%%%%%%%%%%%%%%%%%%%%%%%%%%%%%%%%%%%%%%%%%%%%%
%%%%%%%%%%%%%%%%%%%%%%%%%%%%%%%%%%%%%%%%%%%%%%%%%%%%%%%%%%%%%%%%%%%%%%%%%%%%%%%%%%%%%%%%
\acknowledgments
%%%%%%%%%%%%%%%%%%%%%%%%%%%%%%%%%%%%%%%%%%%%%%%%%%%%%%%%%%%%%%%%%%%%%%%%%%%%%%%%%%%%%%%%
This work was supported by the Estonian Research Council through grants PRG1055, PRG1677, RVTT3, RVTT7, TARISTUS24-TK10, TARISTUS25-TK3 and the CoE program TK202 ``Foundations of the Universe''. This article is based on work carried out within COST Actions COSMIC WISPers CA21106 and CosmoVerse CA21136, supported by COST (European Cooperation in Science and Technology).

\bibliographystyle{JHEP}
\bibliography{references}

\end{document}